\documentclass[longbibliography,physrev,aps,twocolumn]{revtex4-2}
\usepackage{graphicx,epsfig}
\usepackage{times}
\usepackage{amsmath,amssymb,wasysym}
\usepackage{dsfont} % \mathds
\usepackage{color}
\usepackage[linktocpage=true,colorlinks=true, pdfborder={0 0 0},linkcolor=blue,citecolor=blue,urlcolor=blue,anchorcolor=black,linkcolor=blue]{hyperref}

\DeclareMathOperator{\rank}{rank}

\newcommand{\op}[1]{\hat{#1}}

\newcommand{\normV}[1]{||{#1}||_2}
\newcommand{\normM}[1]{||{#1}||}
\newcommand{\normspec}[1]{||{#1}||_{\text{2}}}

\newcommand{\ident}{\mathds{1}}
\newcommand{\C}{\mathbb{C}}

\newcommand{\R}{\mathbb{R}}

\newcommand{\quoting}[1]{``#1''}

\newcommand{\rem}[1]{}

\newcommand{\imagc}[1]{\text{Im}\,#1}

\newcommand{\realc}[1]{\text{Re}\,#1}
\newcommand{\oHami}{\op{\Gamma}}
\newcommand{\Hilbert}{{\cal H}}

\newcommand{\ket}[1]{|#1\rangle}
\newcommand{\transpose}{{\text{T}}}

\newcommand{\braket}[2]{\langle#1|#2\rangle}

\newcommand{\sandwich}[3]{\langle#1|#2|#3\rangle}
\newcommand{\prodFro}[2]{\langle#1,#2\rangle_{\text{F}}}
\newcommand{\normFro}[1]{||#1||_{\text{F}}}
\newcommand{\rca}{\xi}
\newcommand{\rcamax}{\rca_{\text{ub}}}
\newcommand{\rcb}{\zeta}
\newcommand{\pseudospectrum}{\Lambda}

\DeclareMathOperator{\trace}{Tr}

\newcommand{\Hs}{\op{H}_0}
\newcommand{\Hp}{\op{H}_1}
\newcommand{\Ha}{\op{H}}

\newcommand{\pf}{\omega}
\newcommand{\order}{n}

\newcommand{\ev}{E}
\newcommand{\evEP}{\ev_{\text{EP}}}
\newcommand{\freqEP}{\omega_{\text{EP}}}
\newcommand{\state}{\psi}
\newcommand{\stateEP}{\state_{\text{EP}}}
\newcommand{\Jordan}{j}
\newcommand{\GF}{\op{G}}
\newcommand{\Gn}{\op{G}}
\newcommand{\svalue}{\sigma}
\newcommand{\svaluemax}{\sigma_1}
\newcommand{\pseps}{\tilde{\varepsilon}}

\begin{document}

\title{Response strengths of open systems at exceptional points}
\author{Jan Wiersig}
\affiliation{Institut f{\"u}r Physik, Otto-von-Guericke-Universit{\"a}t Magdeburg, Postfach 4120, D-39016 Magdeburg, Germany}
\email{jan.wiersig@ovgu.de}
\date{\today}
\begin{abstract}
Open quantum and wave systems exhibit exotic degeneracies at exceptional points in parameter space that have attracted considerable attention in various fields of physics, including optics and photonics. One reason is the strong response of open systems at such degeneracies to external perturbations and excitations. We introduce two characteristics of exceptional points that quantify the response in terms of energy eigenvalues and eigenstates, intensity, and dynamics. The concept is verified for several physically relevant examples. This work provides a new perspective on the physics of exceptional points. 

\end{abstract}
%\pacs{42.25.-p, 42.60.Da, 42.25.Dd}
% 42.25.-p Wave optics, includes:
%          42.25.Gy Edge and boundary effects; reflection and refraction
%          42.25.Fx Diffraction and scattering
%          42.25.Hz Interference
% 42.55.Sa Microcavity and microdisk lasers
% 42.60.Da Resonators, cavities, amplifiers, arrays, and rings
% 42.25.Dd Wave propagation in random media
% 05.45.Mt Quantum chaos; semiclassical methods
\maketitle

\section{Introduction}
\label{sec:intro}
% general
% non-Hermitian Hamiltonian
Various aspects of open systems are often very well described by a non-Hermitian effective Hamiltonian~\cite{Feshbach58,Feshbach62}, e.g., atoms in optical potentials~\cite{KOA97,BerryODell98}, microwave cavities~\cite{SPK02}, one-dimensional nanostructures~\cite{CK09}, nuclear physics~\cite{MRW10}, optical microcavities~\cite{Wiersig11,KYW18,KW19}, parity-time-symmetric electronics~\cite{SLL12}, and coupled laser arrays~\cite{DM19}.
% consequences
The non-Hermiticity or non-self-adjointness of the Hamiltonian $\Hs \neq \Hs^\dagger$ implies that the energy eigenvalues can be complex-valued with the imaginary part determining a decay or growth rate. Furthermore, if the Hamiltonian is non-normal, meaning that $[\Hs,\Hs^\dagger] \neq 0$, then its (right) eigenstates are in general mutually nonorthogonal. The nonorthogonality can be strong near exceptional points (EPs) in parameter space~\cite{Kato66,Heiss00,Berry04,Heiss04,MA19}, where at least two eigenstates become collinear. 
% existence 
EPs have been observed in numerous experiments, e.g., in microwave cavities~\cite{DGH01,DDG04,DFM07,WHL19}, optical microcavities~\cite{LYM09,POL14,POL16,RZZ19}, coupled atom-cavity composites~\cite{CKL10}, photonic lattices~\cite{RBM12}, semiconductor exciton-polariton systems~\cite{GEB15}, and ultrasonic cavities~\cite{SKM16}. 

% example
A simple and informative example of a linear operator at an EP is the $2\times 2$ matrix
\begin{equation}\label{eq:H0EP}
\Hs = \left(\begin{array}{cc}
E_0 & A_0 \\
0 & E_0\\
\end{array}\right) \ ,
\end{equation}
which for $A_0\neq 0$ has only one eigenvector $(1,0)^\transpose$, so the geometric multiplicity is 1;  the superscript $^\transpose$ marks the transpose of the vector. Its eigenvalue is $\evEP = E_0\in\C$ with an algebraic multiplicity of 2. The special case $A_0 = 0$ is a conventional degeneracy, also called diabolic point (DP)~\cite{BW84}, where only the eigenvalues degenerate but two orthogonal eigenvectors can be chosen, e.g., $(1,0)^\transpose$ and $(0,1)^\transpose$. Here, both the geometric and the algebraic multiplicity are 2.

% Order of EP
One characteristic property of an EP is its order~$\order \geq 2$. At an EP of order~$\order$ (EP$_\order$) exactly $\order$ eigenvalues and the corresponding eigenstates coalesce. The geometric multiplicity is 1 while the algebraic multiplicity is~$\order$.
% EP-sensitivity
When a non-normal Hamiltonian $\Hs$ at an EP$_\order$ is subjected to a small perturbation of non-dimensional strength~$\varepsilon > 0$,
\begin{equation}\label{eq:H}
\Ha = \Hs+\varepsilon\Hp \ ,
\end{equation}
then the resulting energy (or frequency) splittings are generically proportional to the $\order$th root of $\varepsilon$~\cite{Kato66}, which for sufficiently small perturbations is larger than the linear scaling near a DP. 
Shortly after it had been suggested to exploit these larger splittings at EPs for sensing applications~\cite{Wiersig14b}, several experiments have proven the feasibility of EP-based sensors~\cite{COZ17,HHW17,ZCZ18,CSH18,DLY19,ZSL19,LLS19,HSC19,PNC20}; a review of recent progress can be found in Ref.~\cite{Wiersig20c}.
% example
To continue the example in Eq.~(\ref{eq:H0EP}) we add the perturbation Hamiltonian 
\begin{equation}\label{eq:H1}
\Hp = \left(\begin{array}{cc}
C_1 & A_1 \\
B_1 & C_1\
\end{array}\right)
\end{equation}
to $\Hs$ as in Eq.~(\ref{eq:H}) leading to the change of energy eigenvalues
\begin{equation}\label{eq:EPsplitting}
\ev_j-\evEP = \sqrt{\varepsilon}\sqrt{A_0B_1} + {\cal O}(\varepsilon) \ ,
\end{equation}
where the positive sign of the square root corresponds to $\ev_1$ and the negative to $\ev_2$. The splitting between these two eigenvalues is proportional to $\sqrt{\varepsilon}$ for small $\varepsilon$, as expected for an EP of second order. Notice that~$A_0$ is the only parameter of the unperturbed Hamiltonian~(\ref{eq:H0EP}) that determines the size of the splitting. This quantity has therefore been called the \quoting{strength} of the EP~\cite{Wiersig18b}.

% purpose of the present paper
The aim of the present paper is to generalize the concept of the \quoting{strength} of an EP to $\order\times\order$ Hamiltonians at an EP$_\order$, where the order~$\order$ can be two or larger. For this we introduce two {\em response strengths} which describe the response of a given quantum system at an EP in at least four different situations: (i) the spectral response to perturbations, (ii) the eigenstate response to perturbations, (iii) the intensity response to excitations, and (iv) the dynamic response to initial deviations from the EP eigenstate.

% outline of the paper
The outline of the paper is as follows. Section~\ref{sec:MP} introduces some necessary mathematical preliminaries. In Sec.~\ref{sec:RP} we discuss the response of quantum systems at an EP to perturbations. In Sec.~\ref{sec:splitting} we derive an upper bound for the change of the eigenenergies and introduce the spectral response strength. The relation to pseudospectra is covered in Sec.~\ref{sec:pseudospectra}. An upper bound for this response strength in passive systems is derived in Sec.~\ref{sec:passive}. In Sec.~\ref{sec:eigenstates} we show that the quantification of the eigenstate response to perturbations requires a second response strength. Sections~\ref{sec:rtoe} and~\ref{sec:dynamic} reveal that the spectral response strength also quantifies the intensity response to excitations and a certain type of dynamic response. Section~\ref{sec:examples} illustrates the concept for a number of physically relevant examples. A summary is given in Sec.~\ref{sec:summary}.

\section{Mathematical preliminaries}
\label{sec:MP}
This section aims at giving a basic introduction to some mathematical concepts to set the background for this article. 

\subsection{Matrix norms}
\label{sec:norms}
Physicists are familiar with vector norms, or more precisely with the vector 2-norm $\normV{\psi} = \sqrt{\braket{\psi}{\psi}}$ of a vector $\ket{\psi}$ based on the usual inner product in complex vector space. Matrix norms are much less familiar to physicists. A norm $\normM{\op{A}}$ is a mapping of a matrix $\op{A}$ to the nonnegative real numbers. The defining properties of the matrix norm, see e.g. Ref.~\cite{HJ13}, are for all matrices $\op{A}$, $\op{B}$ and complex numbers $\alpha$ (i) $\normM{\op{A}+\op{B}}\leq\normM{\op{A}}+\normM{\op{B}}$, (ii) $\normM{\alpha \op{A}} = |\alpha|\normM{\op{A}}$, (iii) $\normM{\op{A}} = 0$ if and only if $\op{A}=0$, and (iv) $\normM{\op{A}\op{B}} \leq \normM{\op{A}}\,\normM{\op{B}}$. The nonnegativity of $\normM{\op{A}}$ follows from (i) and (ii).

On top of these defining properties we require the matrix norm to be compatible with the vector 2-norm and to be unitarily invariant. A matrix norm is said to be compatible with the vector 2-norm if for all matrices $\op{A}$ and vectors $\ket{\psi}$ 
\begin{equation}\label{eq:compatible}
\normV{\op{A}\psi} \leq \normM{\op{A}}\,\normV{\psi} \ .
\end{equation}
A matrix norm is said to be unitarily invariant if for all matrices $\op{A}$ and all unitary matrices~$\op{U}$ and~$\op{V}$
\begin{equation}\label{eq:unitary}
\normM{\op{U}\op{A}\op{V}} = \normM{\op{A}} \ .
\end{equation}
Prominent examples of matrix norms fulfilling Eqs.~(\ref{eq:compatible}) and~(\ref{eq:unitary}) are the Frobenius norm
\begin{equation}\label{eq:Fronorm}
\normFro{\op{A}} := \sqrt{\trace{(\op{A}^\dagger\op{A})}} 
\end{equation}
with the trace $\trace$ and the spectral norm
\begin{equation}\label{eq:defspn}
\normspec{\op{A}} := \max_{\normV{\psi} = 1}\normV{\op{A}\psi} \ .
\end{equation}
As usual, the same notation is used for the spectral norm and the vector 2-norm, but the meaning is always clear from the context.
Both the Frobenius and the spectral norm can be expressed by the singular values $\svalue_1 \geq \svalue_2 \geq \ldots \geq 0$ of the considered matrix $\op{A}$ as
\begin{equation}
\label{eq:normsv}
\normFro{\op{A}} = \sqrt{\sum_j\svalue_j^2} 
\;\;\text{and}\;\;
\normspec{\op{A}} = \svaluemax \ .
\end{equation}
From the last relations it is evident that
\begin{equation}\label{eq:sleqF}
\normspec{\op{A}} \leq \normFro{\op{A}} \ ,
\end{equation}
with the equality for rank-1 matrices, since the rank of a matrix is equal to the number of non-zero singular values.
In addition, the spectral norm is even the smallest norm that is compatible with the vector 2-norm. This beneficial statement follows by a comparison of Eqs.~(\ref{eq:compatible}) and~(\ref{eq:defspn}).

\subsection{Jordan vectors}
\label{sec:jv}
We consider in this article an $\order\times\order$-Hamiltonian $\Hs$ at an EP of order $\order$ with eigenvalue $\evEP\in\C$ and eigenstate $\ket{\stateEP}$. To establish a basis in the $\order$-dimensional generalized eigenspace $\order-1$ more vectors are needed. The basis can be provided by the linearly independent Jordan vectors $\ket{\Jordan_1}, \ket{\Jordan_2}, \ldots, \ket{\Jordan_\order}$ defined by the Jordan chain (see, e.g., Ref.~\cite{SM03})
\begin{eqnarray}\label{eq:jc1}
(\Hs-\evEP\ident)\ket{\Jordan_1} & = & 0 \ ,\\ 
\label{eq:jc}
(\Hs-\evEP\ident)\ket{\Jordan_l} & = & \ket{\Jordan_{l-1}} 
\;;\, l = 2,\ldots,\order \ ,
\end{eqnarray}
where $\ident$ denotes the identity. Only $\ket{\Jordan_1}$ is an eigenstate of the Hamiltonian, which is $\ket{\Jordan_1} = \ket{\stateEP}$. The other Jordan vectors are often called generalized eigenvectors. These vectors are not uniquely determined and can be redefined as
\begin{eqnarray}\label{eq:altJv}
  \ket{\Jordan_1'} & = & c_1\ket{\Jordan_1}\\
  \label{eq:altJv2}
\ket{\Jordan_2'} & = & c_1\ket{\Jordan_2} + c_2\ket{\Jordan_1}\\
\nonumber
&\vdots&\\
 \label{eq:altJvn}
\ket{\Jordan_{\order}'} & = & c_1\ket{\Jordan_{\order}} + c_2\ket{\Jordan_{\order-1}} + \ldots + c_{\order}\ket{\Jordan_1}
\end{eqnarray}
where $c_1,\ldots,c_\order$ are arbitrary complex numbers with $c_1 \neq 0$. In the following we use this freedom to require 
\begin{eqnarray}\label{eq:ortho1}
\braket{\Jordan_1}{\Jordan_1} & = & 1 \ ,\\
\label{eq:ortho2}
\braket{\Jordan_\order}{\Jordan_l} & = & 0
\quad\mbox{for}\; l = 1, \ldots, \order-1 \ ,
\end{eqnarray}
where here and henceforth we suppress the $'$.

%\newpage
\section{Response to perturbations}
\label{sec:RP}
\subsection{Spectral response}
\label{sec:splitting}
In this section we consider the response of a system at an EP to perturbations. Similar concepts can be found in the mathematical literature, e.g., in Refs.~\cite{Sluis75,Cho97ifa,Karow06}. However, our approach is more elementary and presented in physical terms, such as the Green's function, which is useful also for later sections. Our starting point is the eigenvalue equation of the Hamiltonian~(\ref{eq:H})
\begin{equation}\label{eq:EP}
(\Hs+\varepsilon\Hp)\ket{\state_j} = \ev_j\ket{\state_j}
\end{equation}
with eigenvalues $\ev_j$ and eigenvectors $\ket{\state_j}$ normalized to unity, i.e., $\normV{\state_j} = 1$. Equation~(\ref{eq:EP}) can be written as
\begin{equation}\label{eq:EP2}
\ket{\state_j} = \varepsilon\GF(\ev_j)\Hp\ket{\state_j}
\end{equation}
with the Green's function (resolvent) of the unperturbed Hamiltonian 
\begin{equation}\label{eq:GFdef}
\GF(\ev) := (\ev\ident-\Hs)^{-1} \ . 
\end{equation}
Taking the vector norm on both sides of Eq.~(\ref{eq:EP2}) and using the normalization of the eigenvector gives
\begin{equation}\label{eq:EP3}
1 = \varepsilon\normV{\GF(\ev_j)\Hp\state_j} \ .
\end{equation}
Exploiting the compatibility of the matrix norm to the vector norm~(\ref{eq:compatible}) twice we get the inequality
\begin{equation}\label{eq:EPineq}
1 \leq \varepsilon\normM{\GF(\ev_j)}\,\normM{\Hp} \ .
\end{equation}

In the following, we consider the $\order\times\order$ Hamiltonian $\Hs$ to be at an EP$_\order$ with eigenvalue $\evEP$ and restrict ourselves to an $\order$-dimensional Hilbert space~$\Hilbert$. In this generalized eigenspace the matrix
\begin{equation}\label{eq:N}
\op{N} := \Hs-\evEP\ident 
\end{equation}
is nilpotent of index $\order$; hence, $\op{N}^\order = 0$ but $\op{N}^{\order-1} \neq 0$.

% Heiss
According to Refs.~\cite{Kato66,Heiss15} the Green's function near an EP$_\order$ is
\begin{equation}\label{eq:Heiss}
\GF(\ev) = \frac{\ident}{\ev-\evEP} + \frac{\Gn_2}{(\ev-\evEP)^2} + \ldots + \frac{\Gn_\order}{(\ev-\evEP)^\order}
\end{equation}
with the definition
\begin{equation}\label{eq:Mk}
\Gn_k := \op{N}^{k-1} \ . 
\end{equation}
For the convenience of the reader, an elegant derivation is included in Appendix~\ref{app:GF}.
We consider only the $\order$th-order contribution of the Green's function which is the dominant contribution if (i) the energy $\ev_j$ is sufficiently close to $\evEP$ for small perturbation strength $\varepsilon$ and (ii) the generic situation with [cf. Eq.~(\ref{eq:EP3})] 
\begin{equation}
\Gn_\order\Hp\ket{\state_j} \neq 0
\end{equation}
applies. Plugging this leading-order contribution into the inequality~(\ref{eq:EPineq}) gives 
\begin{equation}\label{eq:precentral}
|\ev_j-\evEP|^\order \leq \varepsilon \normM{\Hp}\, \normM{\Gn_\order}  \ .
\end{equation}
This inequality gives an upper bound for the energy splitting $|\ev_j-\evEP|$ near an EP$_\order$. 
% relation to math literature
There is a large body of literature in mathematics on perturbation bounds for eigenvalues of non-normal matrices. In the nondegenerate case this is the celebrated Bauer-Fike theorem~\cite{BF60}, which has the same structure as Eq.~(\ref{eq:precentral}) except for the exponent $\order$ and $\normM{\Gn_\order}$ replaced by the spectral condition number. For the degenerate case, i.e., at an EP, most of the works provide bounds in terms of the eigenvectors~\cite{MBO97}, with some exceptions, e.g., Refs.~\cite{Sluis75,Cho97ifa}. 
In the later references similar inequalities as~(\ref{eq:precentral}) had been derived. However, their results are less explicit and more difficult to handle than the one presented here.

% separation of bound -> rca
The bound given by Eq.~(\ref{eq:precentral}) can be seen also as the spectral radius of $\Ha-\evEP\ident$. It factorizes into two parts, the first part depends only on the perturbation strength~$\varepsilon$ and the perturbation Hamiltonian~$\Hp$. In contrast, the second part depends via Eqs.~(\ref{eq:N}) and (\ref{eq:Mk}) only on the unperturbed Hamiltonian. We call it the spectral response strength associated to the EP
\begin{equation}\label{eq:rca}
\rca := \normspec{\Gn_\order} \ .
\end{equation}
Here, we have specified the norm to be the spectral norm. The spectral norm is privileged because it is the smallest norm compatible with the vector 2-norm. It therefore gives the tightest bounds. Inequality~(\ref{eq:precentral}) turns into the important result 
\begin{equation}\label{eq:specresponse}
|\ev_j-\evEP|^\order \leq \varepsilon \normspec{\Hp}\,\rca  \ .
\end{equation}
% similarity transformation
As the spectral norm is unitarily invariant~(\ref{eq:unitary}), the nonnegative number $\rca$ is invariant under a unitary similarity transformation of $\Hs$. This is an indispensable property as $\rca$ and the right-hand side (RHS) of the inequality~(\ref{eq:specresponse}) should be unaffected by a unitary change of basis.
Moreover, if one were misapplying Eq.~(\ref{eq:rca}) to a DP$_\order$ then the unitarily invariance would ensure the correct result $\rca = 0$. For an EP, $\rca$ is always larger than zero.

The unit of $\rca$ is Joule$^{\order-1}$ for quantum problems and 1/s$^{\order-1}$ for optical problems (where $\hbar = 1$). It is therefore clear that one should not compare the spectral response strengths for EPs of different order.

% spectral norm = Frobenius norm for Mn
Even though the spectral norm and the Frobenius norm give different values for a general matrix $\op{A}$, see Eq.~(\ref{eq:sleqF}), in the case of $\op{A} = \Gn_\order$ they give exactly the same values, i.e.,
\begin{equation}\label{eq:spMFn}
\normspec{\Gn_\order} = \normFro{\Gn_\order} \ .
\end{equation}
To see this we first note that $\op{N}$ is a nilpotent matrix with index of nilpotency $\order$ and one-dimensional kernel. Hence, the dimension of the kernel of $\op{N}^{\order-1}$ is $\order-1$. As a consequence, $\Gn_\order = \op{N}^{\order-1}$ has rank 1 and therefore only one nonzero singular value. With Eqs.~(\ref{eq:normsv}) this proves Eq.~(\ref{eq:spMFn}). 
We conclude that the response strength $\rca$ can be computed with both the spectral norm and the Frobenius norm leading to exactly the same numerical values. This is a valuable observation as the Frobenius norm is much easier to compute, see Eq.~(\ref{eq:Fronorm}).

Note that $\normspec{\Hp} \neq \normFro{\Hp}$ in general. Thus for $\Hp$ in Eq.~(\ref{eq:specresponse}) the spectral norm has to be used in order to get the tightest bound. However, the philosophy here is not to compute the spectral response for given perturbation $\Hp$ but to quantify the response strength of the system at the EP to generic perturbations. For this purpose it is sufficient to employ the Frobenius norm.

% order neq matrix dimension
Finally it is remarked that the presented theory works only if the order of the EP, $\order$, and the dimension $d$ of the Hilbert space~$\Hilbert$ (and hence the matrix dimensions $d\times d$ of the Hamiltonian) are equal. If $d > \order$ one first has to project the vectors in ~$\Hilbert$ onto the relevant subspace of dimension~$\order$, thereby reducing the Hamiltonian to an $\order\times\order$ matrix. 

\subsection{Relation to pseudospectra}
\label{sec:pseudospectra}
In this subsection we connect the response strength~$\rca$ to the notion of pseudospectra. Pseudospectra are a tool to describe the behavior of non-normal matrices subjected to perturbations. A comprehensive description and overview can be found in the classical monograph~\cite{TE05}. 
% definition of pseudospectrum
Given a positive number $\pseps$, the $\pseps$-pseudospectrum of a non-normal matrix $\Hs$ can be defined as the subset of the complex plane
\begin{equation}\label{eq:ps1}
\pseudospectrum_{\pseps} := \{\ev\in\C: \normM{(E\ident-\Hs)^{-1}} > 1/\pseps\}
\end{equation}
with arbitrary matrix norm $\normM{\cdot}$. The pseudospectrum of $\Hs$ contains the spectrum of $\Hs$, as $\normM{(E\ident-\Hs)^{-1}}$ diverges at the eigenvalues of $\Hs$. 
A recent application to optical systems is discussed in Ref.~\cite{ZE19}. There are other, equivalent ways to introduce the pseudospectrum~\cite{TE05} which have also been applied to optical systems~\cite{MGT14,Makris21}.

% relation to sigma_1
As in the previous subsection we consider an EP of order~$\order$, use the Green's function~(\ref{eq:GFdef}) and its expansion at the EP in Eq.~(\ref{eq:Heiss}), and restrict ourselves to the leading-order contribution. Choosing the spectral norm and applying the definition of $\rca$ in Eq.~(\ref{eq:rca}) we arrive at
\begin{equation}\label{eq:ps2}
\pseudospectrum_{\pseps} = \{\ev\in\C: \sqrt[\order]{\pseps\rca} > |\ev-\evEP|\} \ .
\end{equation}
Hence, the response strength $\rca$ determines the radius~$\sqrt[\order]{\pseps\rca}$ of the pseudospectrum-disk (the so-called $\pseps$-pseudospectral radius~\cite{TE05}) around the EP of order $\order$, which is equivalent to inequality~(\ref{eq:specresponse}) if $\pseps$ is identified with $\varepsilon \normspec{\Hp}$. In this case, our definition~(\ref{eq:rca}) agrees with the H{\"o}lder condition number of the eigenvalue~$\evEP$~\cite{Karow06}.
% The larger $\rca$ the larger the pseudospectrum around the EP is. The latter reflects an enhanced sensitivity to perturbations.

\subsection{Passive systems}
\label{sec:passive}
In this subsection we reveal the existence of an upper bound for~$\rca$ in passive systems. Physically, a system is passive if there is no gain. Mathematically, it means that the decay operator
\begin{equation}\label{eq:decayop}
\oHami := i(\Hs-\Hs^\dagger) 
\end{equation}
is not only Hermitian but also positive semidefinite, see, e.g., Ref.~\cite{Wiersig16} (for caveats see Ref.~\cite{Wiersig19}). We write this equation together with Eq.~(\ref{eq:N}) as 
\begin{equation}\label{eq:oHamiNN}
\op{N}-\op{N}^\dagger = -i(\oHami-\beta\ident) 
\end{equation}
with the definition $\beta := -2\imagc{\evEP}$. As nilpotent matrices both $\op{N}$ and $\op{N}^\dagger$ are traceless~\cite{HJ13}. Therefore,
\begin{equation}\label{eq:troHami}
\trace{\oHami} = \order\beta \ ,
\end{equation}
where again $\order$ is the order of the EP. For a positive semidefinite matrix $\trace{\oHami}\geq 0$ and hence $\beta \geq 0$.

We first deal with an EP$_2$. We square both sides of Eq.~(\ref{eq:oHamiNN}), employ $\op{N}^2 = 0 = \op{N}^{\dagger 2}$ and the cyclic property of the trace to obtain
\begin{equation}\label{eq:NN2}
\trace{(\op{N}^\dagger\op{N})} = \frac{1}{2}\trace{\left[(\oHami-\beta\ident)^2\right]} \ . 
\end{equation}
The left-hand side (LHS) is, for $\order = 2$ according to Eqs.~(\ref{eq:rca}) and~(\ref{eq:spMFn}), the response strength~$\rca$ squared. The positive semidefiniteness of~$\oHami$ imposes an upper bound to the RHS of Eq.~(\ref{eq:NN2}). The maximum value is given when $\oHami$ is a rank-1 matrix. In this case, all eigenvalues of $\oHami$ are zero except one, which, according to Eq.~(\ref{eq:troHami}), assumes the value $2\beta$. The RHS of Eq.~(\ref{eq:NN2}) is $\frac12(\beta^2+\beta^2) = \beta^2$. We conclude that $\rca \leq \rcamax$ with the upper bound 
\begin{equation}\label{eq:passiveEP2}
\rcamax = 2|\imagc{\evEP}| \ .
\end{equation}
This upper bound is a restriction for the spectral response strength of passive systems at an EP$_2$ and therefore for the size of the energy splitting under perturbation. Having in mind that $2|\imagc{\evEP}|$ is the linewidth of the spectral peak at the EP it becomes clear that Eq.~(\ref{eq:passiveEP2}) represents a limitation for the resolvability of the splitting. This is of particular importance for EP-based sensors.

We remark that $\rca = \rcamax$ when the rank of $\oHami$, which can be interpreted as the number of available decay channels, is unity. Examples of single-decay-channel systems are tight-binding chains with a single lossy site~\cite{Wiersig18a} and doorway states in nuclear physics~\cite{SRS97}. 

% EP3
Next, we study the case of an EP of the third order, $\order = 3$. Here, we take the fourth power of both sides of Eq.~(\ref{eq:oHamiNN}), exploit $\op{N}^3 = 0 = \op{N}^{\dagger 3}$ and the cyclic property of the trace to obtain
\begin{equation}\label{eq:NN3}
4\trace{(\op{N}^{\dagger 2}\op{N}^2)} + 2\trace{(\op{N}^\dagger\op{N}\op{N}^\dagger\op{N})} = \trace{\left[(\oHami-\beta\ident)^4\right]} \ . 
\end{equation}
Again, the RHS is bounded from above due to the positive semidefiniteness of~$\oHami$, and the maximum value is approached when $\oHami$ is a rank-1 matrix. In this case, all eigenvalues of $\oHami$ are zero except one, which, according to Eq.~(\ref{eq:troHami}), attains the value $3\beta$. As a consequence, the RHS of Eq.~(\ref{eq:NN3}) is $(16+1+1)\beta^4 = 18\beta^4$. For the LHS we exploit the inequality
\begin{equation}\label{eq:ineqNN}
\trace{(\op{N}^\dagger\op{N}\op{N}^\dagger\op{N})} \geq \trace{(\op{N}^{\dagger 2}\op{N}^2)} \ .
\end{equation}
This inequality can be proven by inserting the two auxiliary matrices $\op{X} := \op{N}^\dagger\op{N}$ and $\op{Y} := \op{N}\op{N}^\dagger$ into the Frobenius inner product~\cite{HJ13}
\begin{equation}\label{eq:Fro}
\prodFro{\op{A}}{\op{B}} := \trace{(\op{A}^\dagger\op{B})} 
\end{equation}
leading to $\prodFro{\op{X}}{\op{Y}} = \trace{(\op{N}^{\dagger 2}\op{N}^2)}$ and $\prodFro{\op{X}}{\op{X}} = \trace{(\op{N}^\dagger\op{N}\op{N}^\dagger\op{N})} = \prodFro{\op{Y}}{\op{Y}}$; the last step takes advantage of the cyclic property of the trace. From the Cauchy-Schwarz inequality
\begin{equation}\label{eq:CSinequality}
|\prodFro{\op{A}}{\op{B}}|^2 \leq \prodFro{\op{A}}{\op{A}}\prodFro{\op{B}}{\op{B}} 
\end{equation}
follows the inequality~(\ref{eq:ineqNN}), which we use to recast Eq.~(\ref{eq:NN3}) as
\begin{equation}
6\trace{(\op{N}^{\dagger 2}\op{N}^2)} \leq \trace{\left[(\oHami-\beta\ident)^4\right]} \leq 18\beta^4 \ .
\end{equation}
With Eqs.~(\ref{eq:rca}) and~(\ref{eq:spMFn}) we reach the result
\begin{equation}\label{eq:passiveEP3}
\rcamax = 4\sqrt{3}|\imagc{\evEP}|^2 \ .
\end{equation}
This upper bound holds for passive systems at an EP$_3$. The additional step involving the inequality~(\ref{eq:ineqNN}) may cause $\rcamax$ to be a nonsharp upper bound. This guess will be proven correct later in Sec.~\ref{sec:examples}.
 
\subsection{Eigenstate response}
\label{sec:eigenstates}
Here, we derive a response strength that describes the response of the eigenstates to perturbations. More precisely, we are interested in the resulting component $\ket{\Delta\state_j}$ orthogonal to the original vector $\ket{\stateEP}$. Hence, the ansatz for the eigenstates of the perturbed Hamiltonian is 
\begin{equation}\label{eq:dv}
\ket{\state_j} = \ket{\stateEP} + \ket{\Delta\state_j} 
\end{equation}
with  
\begin{equation}\label{eq:dvnorm}
\braket{\stateEP}{\Delta\state_j} = 0
\end{equation}
and $\normV{\Delta\state_j}$ small. We write 
\begin{equation}\label{eq:de}
\ev_j = \evEP + \sqrt[\order]{\varepsilon}\Delta\ev_j \ .
\end{equation}
The ansatz for the eigenstates~(\ref{eq:dv}) and eigenvalues~(\ref{eq:de}) is plugged into the eigenvalue equation~(\ref{eq:EP}) of the Hamiltonian~$\Ha$. We use the eigenvalue equation of the unperturbed Hamiltonian $\Hs\ket{\stateEP} = \evEP\ket{\stateEP}$ and keep only the lowest-order terms leading to
\begin{equation}
\op{N}\ket{\Delta\state_j} = \sqrt[\order]{\varepsilon}\Delta\ev_j\ket{\stateEP} 
\end{equation}
where we have exploited Eq.~(\ref{eq:N}). Comparison to Eqs.~(\ref{eq:jc1}) and~(\ref{eq:jc}) reveals that 
\begin{equation}\label{eq:Dvj2}
\ket{\Delta\state_j} = \sqrt[\order]{\varepsilon}\Delta\ev_j\ket{\Jordan_2} + c\ket{\Jordan_1}
\end{equation}
with the Jordan vectors $\ket{\Jordan_1} = \ket{\stateEP}$ and $\ket{\Jordan_2}$. The constant~$c$ has to be adjusted such that condition~(\ref{eq:dvnorm}) is fulfilled. This gives $c = -\sqrt[\order]{\varepsilon}\Delta\ev_j\braket{\Jordan_1}{\Jordan_2}$. Note that for $\order > 2$ the constant $c$ does normally not vanish because of the chosen normalization and orthogonalization conditions~(\ref{eq:ortho1})-(\ref{eq:ortho2}). This is different from other works in the literature~\cite{SM03,Heiss08,DG12}.  

From Eq.~(\ref{eq:Dvj2}) follows
\begin{equation}\label{eq:evev}
\normV{\Delta\state_j} = \sqrt[\order]{\varepsilon}|\Delta\ev_j| \sqrt{\normV{\Jordan_2}^2-|\braket{\Jordan_1}{\Jordan_2}|^2} \ .
\end{equation}
With Eq.~(\ref{eq:specresponse}) we get another important result 
\begin{equation}\label{eq:centralvec}
\normV{\Delta\state_j}^\order \leq \varepsilon \normspec{\Hp}\, \rcb \ , 
\end{equation}
where we have introduced the eigenstate response strength
\begin{equation}\label{eq:rcb}
\rcb := \left({\normV{\Jordan_2}^2-|\braket{\stateEP}{\Jordan_2}|^2}\right)^{\order/2} \rca  \ .
\end{equation}
The nonnegative response strength $\rcb$ describes the response of the eigenstates to perturbations. Its unit is Joule$^{-1}$ (s in optics). The strength $\rcb$ is not only invariant with respect to a unitary similarity transformation of $\Hs$ but also under the redefinition of Jordan vectors in Eqs.~(\ref{eq:altJv})-(\ref{eq:altJv2}).

In the special case $\order = 2$ the two response strengths are not independent. In Appendix~\ref{app:Jordan} it is shown that $\normFro{\Gn_\order} = 1/\normV{j_\order}$. With $\order = 2$ and Eqs.~(\ref{eq:ortho2}), (\ref{eq:spMFn}), and~(\ref{eq:rcb}) it follows
\begin{equation}\label{eq:rcarcb}
\rca\rcb = 1 \ .
\end{equation}
The upper bound of $\rca$ for passive systems discussed in Sec.~\ref{sec:passive} turns to a lower bound of $\rcb$. For $\order > 2$, however, such a conclusion cannot be drawn.

\section{Response to excitations}
\label{sec:rtoe}
So far we have discussed the response to perturbations. In this section, we consider the response to time-harmonic excitations in terms of the norm (intensity in the optical context) of the resulting state. We take into account the inhomogeneous Schr{\"o}dinger equation  
\begin{equation}\label{eq:iSe}
i\hbar\frac{d}{dt}\ket{\psi} = \Hs\ket{\psi} + e^{-i\pf t}P\ket{p}
\end{equation}
with excitation power $P\geq 0$, excitation frequency $\pf\in\R$, and the excitation vector $\ket{p}$ normalized to unity. In quantum mechanics, such an inhomogeneous Schr{\"o}dinger equation may appear unnatural at first sight, but it can be used within the framework of mean-field theory; see, e.g., Ref.~\cite{DCG18}. In classical optics such an equation is often used (with $\hbar = 1$) to describe the excitation of the system by an attached waveguide~\cite{POL16,Sunada18,Langbein18,KW19}.

We are interested here in a steady-state behavior of (the norm of) the state $\ket{\psi}$ in the limit of long times. This long-time limit is in general only finite if all eigenvalues of $\Hs$ have a nonpositive imaginary part. This condition is obviously met by passive systems. But also non-passive systems can fulfill this condition. If so they may have interesting transient dynamics~\cite{MGT14,Makris21}. In the following we assume that the long-time limit of $\ket{\psi}$ is finite. In this case it is given by the particular solution of Eq.~(\ref{eq:iSe}) in terms of the Green's function~(\ref{eq:GFdef})  
\begin{equation}\label{eq:particular}
\ket{\psi(t)} = \GF(\hbar\pf)e^{-i\pf t}P\ket{p} \ .
\end{equation} 
Once more, we consider the Hamiltonian $\Hs$ to be at an EP$_\order$ with eigenvalue $\evEP$. Assuming the generic situation
\begin{equation}\label{eq:intensitygeneric}
\Gn_\order\ket{p} \neq 0
\end{equation}
we restrict ourselves to the $\order$th-order contribution of the Green's function~(\ref{eq:Heiss}) yielding
\begin{equation}
\normV{\psi} = \frac{P}{|\hbar\pf-\evEP|^{\order}} \normV{\Gn_\order p} \ .
\end{equation}
For a matrix norm $\normM{\cdot}$ that is compatible with the vector 2-norm [Eq.~(\ref{eq:compatible})] holds
\begin{equation}\label{eq:compatible1}
\normV{\Gn_\order p} \leq \normM{\Gn_\order}\,\normV{p} \ .
\end{equation}
Choosing again the spectral norm and using the normalization $\normV{p} = 1$ and the definition of $\rca$ in Eq.~(\ref{eq:rca}), it follows for the intensity response the important result
\begin{equation}\label{eq:rtoe}
\normV{\psi} \leq P\, \frac{1}{|\hbar\pf-\evEP|^{\order}}\, \rca\ .
\end{equation}
The first factor depends only on the excitation and is, as expected, proportional to the excitation strength $P$. The second factor depends on both the excitation (frequency $\pf$) and the EP (eigenvalue $\evEP$). Note that the resulting spectral lineshape is not Lorentzian, see, e.g., Refs.~\cite{PZM17,KhW20,SZM22}. The third factor is the spectral response strength associated to the EP, $\rca$. Its appearance might be surprising at first, but note that both the spectral response and the intensity response are mediated by the Green's function of the unperturbed Hamiltonian.
Inequality~(\ref{eq:rtoe}) is invariant under a unitary change of basis and is consistent with the results for special systems at an EP$_2$ in Refs.~\cite{Sunada18,ZOE20}. 

It has to be emphasized that in the derivation of inequality~(\ref{eq:rtoe}) the limitation to the highest-order contribution of the Green's function in the expansion~(\ref{eq:Heiss}) is possibly not sufficient, as $\hbar\pf$ is real and cannot approach the complex energy $\evEP$. In such a situation, the next-order contribution can be significant and the inequality~(\ref{eq:rtoe}) may not be fulfilled. For given eigenvalue~$\evEP$ the best-case scenario is obviously the resonant case, i.e., $\hbar\pf = \realc{\evEP}$.

% noisy exciation
Note that an analogous inequality as~(\ref{eq:rtoe}) can be derived for noisy excitations. The best starting point for such a calculation is the pump operator in Ref.~\cite{Wiersig20}. The calculation for the Frobenius norm is straightforward and not shown here. 

% relation to Petermann factor
It is worth mentioning that $\rca$ in Eq.~(\ref{eq:rtoe}) plays a similar role as the Petermann factor quantifying the mode nonorthogonality in open quantum and wave systems, see, e.g., Ref.~\cite{HS20}. However, there are crucial differences. First, while $\rca$ is defined only at the EP, the Petermann factor is defined everywhere except at the EP, where it diverges to infinity~\cite{Berry03}. Second, $\rca$ describes the $\order$-dimensional generalized eigenspace of the EP, the Petermann factor describes single eigenstates. Third, $\rca$ defines an upper bound for the response, whereas the Petermann factors determine directly, in the context of intensity response, the strength of the response.

We conclude that the spectral response strength also quantifies the intensity response of quantum systems at an EP to harmonic excitations.

\section{Dynamic response to initial deviations from the EP eigenvector}
\label{sec:dynamic}
In this section we show that the long-time response to initial deviations from the EP eigenvector can be expressed by the spectral response strength $\rca$. The Hamiltonian $\Hs$ is again at an EP$_\order$ with eigenvector $\ket{\stateEP}$ and eigenvalue $\evEP$. We examine the non-unitary time-evolution operator $\op{U}(t) = e^{-\frac{i}{\hbar}\Hs t}$ mapping the initial state $\ket{\psi(0)}$ in the generalized eigenspace of the EP to the time-evolved state $\ket{\psi(t)} = \op{U}(t)\ket{\psi(0)}$. 
From the nilpotency of $\op{N}$ follows
\begin{equation}
e^{-\frac{i}{\hbar}\op{N}t} = \sum_{j=0}^{\order-1}\frac{1}{j!}\left(-\frac{i}{\hbar}\op{N}t\right)^{j} \ .
\end{equation}
Using Eq.~(\ref{eq:N}) we can write the time-evolution operator as
\begin{equation}
\op{U}(t) = e^{-i\freqEP t}\sum_{j=0}^{\order-1}\frac{(-i t)^{j}}{j!\hbar^j} \op{N}^{j} 
\end{equation}
with complex frequency $\freqEP := \evEP/\hbar$. For the special initial state $\ket{\psi(0)} = \ket{\stateEP}$ we get $\ket{\psi(t)} = e^{-i\freqEP t}\ket{\stateEP} =: \ket{\stateEP(t)}$. 
For a generic initial state $\ket{\psi(0)} \neq \ket{\stateEP}$ in the generalized eigenspace we can assume 
\begin{equation}
\Gn_\order\ket{\psi(0)} \neq 0 \ . 
\end{equation}
With the definition of $\Gn_k$ in Eq.~(\ref{eq:Mk}), the long-time behavior is given by
\begin{equation}
\op{U}(t) = e^{-i\freqEP t}\frac{(-i t)^{\order-1}}{(\order-1)!\hbar^{\order-1}} \Gn_\order \ .
\end{equation}
Again, we consider a matrix norm $\normM{\cdot}$ that is compatible with the vector 2-norm; see Eq.~(\ref{eq:compatible}). Hence,
\begin{equation}\label{eq:compatible2}
\normV{\psi(t)} \leq \normM{\op{U}(t)}\,\normV{\psi(0)} \ ,
\end{equation}
where the initial state is supposed to be normalized to unity. As before, we confine ourselves to the spectral norm. Recalling the definition of $\rca$ in Eq.~(\ref{eq:rca}) we finally get for large $t$
\begin{equation}\label{eq:dynamics}
\frac{\normV{\psi(t)}}{\normV{\stateEP(t)}} \leq \frac{|t|^{\order-1}}{(\order-1)!\hbar^{\order-1}}\rca \ .
\end{equation}
Hence, also the upper bound of the long-time intensity response to initial deviations from the EP eigenvector is determined by $\rca$.
%That the spectral response strength enters here is reasonable as the time and the spectral domain are related by a Fourier transformation.
Inequality~(\ref{eq:dynamics}) is, together with Eqs.~(\ref{eq:rca}) and (\ref{eq:rcaJv}), consistent with Ref.~\cite{Longhi18b}.

Although this is not directly related to the present work, it is mentioned that the spectral norm and the Frobenius norm had also been used to determine lower bounds for the resources needed for the construction of a given non-unitary time evolution~\cite{UGR12,Uzdin13}. 

\section{Examples}
\label{sec:examples}

% whispering-gallery microcavities with fully asymmetric backscattering
\subsection{Whispering-gallery microcavities with fully asymmetric backscattering}
\begin{figure}[ht]
\includegraphics[width=0.85\columnwidth]{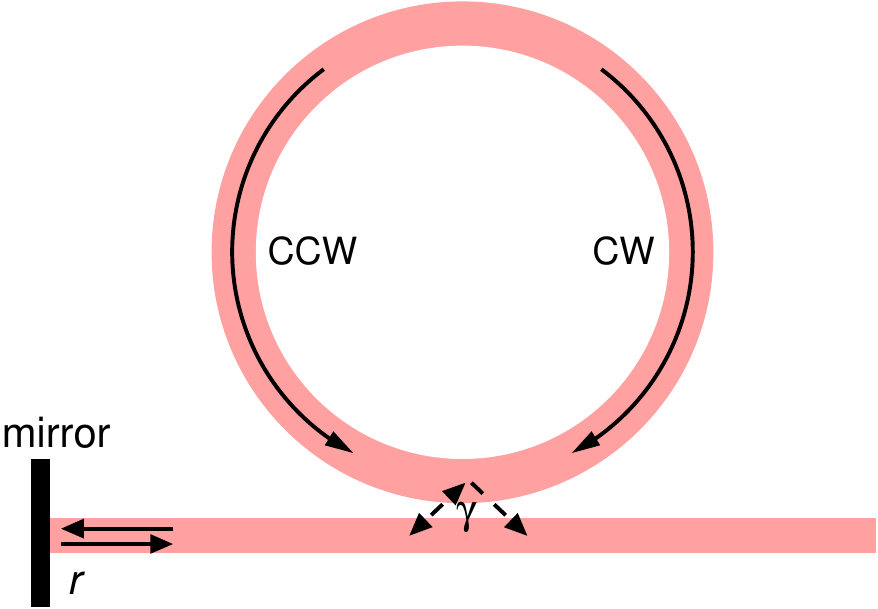}
\caption{Sketch of a microring coupled to a waveguide with coupling rate $\gamma$; see Refs.~\cite{ZRK19,ZOE20}. The semi-infinite waveguide is terminated on the left-hand side by a mirror with field reflection coefficient~$r$. The induced fully asymmetric backscattering of clockwise (CW) and counterclockwise (CCW) propagating waves leads to an EP$_2$.}
\label{fig:asym}
\end{figure}

Our first example is the Hamiltonian~(\ref{eq:H0EP}), which, for instance, describes whispering-gallery microcavities with fully asymmetric backscattering in a two-mode approximation of clockwise (CW) and counterclockwise (CCW) propagating waves~\cite{WKH08,Wiersig11,Wiersig14,POL16,Wiersig18b,ZRK19,ZOE20,QXZ21,SZM22}. In a traveling-wave basis [$(1,0)^\transpose$ for CCW and $(0,1)^\transpose$ for CW], $E_0$ is the frequency of the two propagating waves and $A_0$ is the coefficient for the backscattering of a CW propagating wave into the CCW propagating wave. 
A possible realization introduced in Ref.~\cite{ZRK19} is sketched in Fig.~\ref{fig:asym}. A CW traveling wave that couples to the waveguide with the coupling rate $\gamma \geq 0$ is reflected at the mirror with the field reflection coefficient $r\geq 0$ and can therefore couple back into the microring where it propagates in the CCW direction.  This model has been experimentally realized using a microsphere~\cite{QXZ21,SZM22} and it has been proposed to use this model for EP-based optical amplifiers~\cite{ZOE20}. In the latter work the backscattering coefficient in the Hamiltonian~(\ref{eq:H0EP}) has been calculated to be
\begin{equation}\label{eq:Zhong}
A_0 = -2i\gamma r e^{i\phi}
\end{equation}
where $\phi$ is an additional phase originating from reflection and propagation in the waveguide. There is no backscattering into the other sense of rotation provided that coupling between waveguide and microring is small enough.
For $A_0\neq 0$ the system is at an EP$_2$ as the only eigenvector is $(1,0)^\transpose$. This corresponds to a chiral state, which is a purely CCW propagating wave.
% chirality
Chirality can be seen as another general property of EPs which characterizes a preferred sense of rotation in mode space~\cite{HH01,DDG03} or real space~\cite{WKH08,Wiersig11,Wiersig18b}.

% splitting and spectral response strength
A straightforward calculation of the spectral response strength in Eq.~(\ref{eq:rca}) using Eqs.~(\ref{eq:N}), (\ref{eq:Mk}), and~(\ref{eq:spMFn}) yields 
\begin{equation}\label{eq:wgma}
  \rca = |A_0| \ .
\end{equation}
Hence, the spectral response strength is the absolute value of the  backscattering coefficient. A system at the DP with $A_0 = 0$ has a zero spectral response strength. 
% Ramy's paper
For the particular example in Ref.~\cite{ZOE20} (Fig.~\ref{fig:asym}) we combine Eq.~(\ref{eq:Zhong}) and Eq.~(\ref{eq:wgma}) to find $\rca = 2\gamma r$. Hence, the spectral response strength is expressed by the two physical quantities of the setup, the coupling rate~$\gamma$ and field reflection coefficient~$r$. To increase the response at the EP one therefore has to increase either $r$ or $\gamma$. However, the former cannot be increased beyond $1$ and increasing the latter also broadens the spectral linewidth and at some point may introduce additional backscattering which spoils the fully asymmetric backscattering of CW and CCW propagating waves.

% spectral bound valid?
For the perturbation Hamiltonian~(\ref{eq:H1}) one can confirm by direct calculation that the inequality~(\ref{eq:specresponse}) is indeed fulfilled. The equality holds for $C_1 = 0$ and $|A_1| \leq |B_1|$.
  
% passive system
For the special Hamiltonian~(\ref{eq:H0EP}) it has been shown in Ref.~\cite{Wiersig16} that if the system is passive then
\begin{equation}\label{eq:passive}
|A_0| \leq 2|\imagc{\evEP}| \ ,
\end{equation}
with $\evEP = E_0$. This inequality together with Eq.~(\ref{eq:wgma}) fully agrees with the upper bound $\rcamax$ in Eq.~(\ref{eq:passiveEP2}). 
It is interesting to compare this to the experimental data in Ref.~\cite{COZ17} on an EP$_2$ in a microtoroid perturbed by two nano-tips. Here, the linewidth is $2|\imagc{\evEP}| \approx  2\times 10^7$s$^{-1}$ and the backscattering coefficient is $|A_0| \approx 2\times 10^7$s$^{-1}$. It can be seen that the spectral response strength $\rca = |A_0|$ reaches the upper bound in Eq.~(\ref{eq:passiveEP2}). We conclude that the experiment in Ref.~\cite{COZ17} is already optimized for the passive system under study. It shows the strongest possible backscattering and therefore the strongest possible spectral response strength. The latter can only be further enhanced by introducing optical gain into the system (which has been implemented in Ref.~\cite{COZ17}).

% eigenstate response strength
From Eq.~(\ref{eq:rcarcb}) it can be read that the eigenstate response strength is $\rcb = 1/|A_0|$. This result can be confirmed for this system by an elementary calculation of the eigenvector change perpendicular to the EP eigenvector under the perturbation~(\ref{eq:H1}) in first order
\begin{equation}
  \ket{\Delta\state_j} = \sqrt{\varepsilon}\sqrt{\frac{B_1}{A_0}}\left(\begin{array}{c}
0 \\
1\\
\end{array}\right) \ ,
\end{equation}
where the sign of the square root labels the two vectors. Hence, the change of the eigenstates under perturbation is small if the backscattering coefficient is large. That means a large backscattering coefficient stabilizes the state, in the sense that is more robust against, e.g., fabrication tolerances. This can be of relevance for unidirectional lasing~\cite{POL16} and orbital angular momentum lasing~\cite{MZS16} based on an EP.

The limiting case $A_0\to 0$ is a second-order DP. Here, the eigenstate response strength $\rcb$ diverges. This is analog to the divergences in perturbation theory in quantum mechanics, which can be cured by degenerate-state perturbation theory.

% intensity response
The Green's function~(\ref{eq:GFdef}) of the unperturbed Hamiltonian~(\ref{eq:H0EP}) can be calculated without effort (see also Ref.~\cite{KhW20})
\begin{equation}\label{eq:GFasym1}
\GF(\ev) = \frac{\ident}{\ev-\evEP}+\frac{1}{(\ev-\evEP)^2}
\left(\begin{array}{cc}
0 & A_0 \\
0 & 0\\
\end{array}\right) \ ,
\end{equation}
which is in accordance with the general expansion~(\ref{eq:Heiss}). Obviously, if $A_0 \neq 0$ and $\ev$ approaches $\evEP$ the second term on the RHS is the dominant one. This fact has been exploited in Secs.~\ref{sec:RP} and \ref{sec:dynamic} when we discussed the response to perturbations and the dynamic response. However, for the intensity response to harmonic excitations in Sec.~\ref{sec:rtoe} the energy $\ev = \hbar\pf$ is real-valued and therefore can approach $\evEP$ only in the special case $\imagc{\evEP} = 0$. In the case $\imagc{\evEP} \leq 0$ the closed approach is for $\hbar\pf = \realc{\evEP}$, i.e., on resonance. In this case, Eq.~(\ref{eq:GFasym1}) can be written as
\begin{equation}\label{eq:GFasym2}
\GF(\hbar\pf) = \frac{\ident}{|\imagc\evEP|}+\frac{A_0}{|\imagc\evEP|^2}
\left(\begin{array}{cc}
0 & 1 \\
0 & 0\\
\end{array}\right) \ .
\end{equation}
Hence, the second term is only dominant if $|A_0| \gg |\imagc\evEP|$. But this is out of reach for passive systems, cf. Eq.~(\ref{eq:passive}).

% nongeneric intensity response
Notice that the condition for the generic situation in Eq.~(\ref{eq:intensitygeneric}) can be broken in the case of the Green's function~(\ref{eq:GFasym1}) by exciting only the CCW propagation direction. This is understandable as in this case there is no asymmetric backscattering, so the non-normality of $\Hs$ is not probed.

% time evolution
A short calculation shows that the time evolution operator corresponding to the Hamiltonian~(\ref{eq:H0EP}) is given by
\begin{equation}
\op{U}(t) = e^{-i\freqEP t}\left[\ident -\frac{itA_0}{\hbar}\left(\begin{array}{cc}
0 & 1\\
0 & 0\\
\end{array}\right)\right] \ .
\end{equation}
We can see that the spectral response strength in Eq.~(\ref{eq:wgma}) determines the long-time behavior as summarized in Eq.~(\ref{eq:dynamics}).

% PT-symmetric dimer
\subsection{Parity-time-symmetric dimer}
A further example is the parity-time-symmetric dimer with Hamiltonian
\begin{equation}\label{eq:HsPT}
\Hs = \left(\begin{array}{cc}
\omega_0+i\alpha & g\\
g   & \omega_0-i\alpha\\
\end{array}\right) \ .
\end{equation}
The real-valued quantity $\omega_0$ is the frequency, $\alpha \geq 0$ the gain/loss coefficient, and $g \geq 0$ the coupling strength. 
This system can be realized for instance by two coupled waveguides~\cite{RMG10} or resonators~\cite{HHW17}. It is parity-time symmetric as it is invariant under the combined action of parity (exchange of waveguide/resonators) and time-reversal (exchange of gain and loss) operations; see Ref.~\cite{EMK18} for a recent review on parity-time-symmetric systems.
The eigenvalues of the Hamiltonian~(\ref{eq:HsPT}) are given by (see, e.g., the Methods section of Ref.~\cite{HHW17})
\begin{equation}\label{eq:e1evH}
E_j = \omega_0 \pm \sqrt{g^2-\alpha^2} \ .
\end{equation}
The eigenvalues are degenerate for $g = \alpha$. If $\alpha \neq 0$ this degeneracy is a second-order EP with a single eigenvector $(1,-i)^\transpose/\sqrt{2}$. 
In this case the spectral response strength~(\ref{eq:rca}) is
\begin{equation}
\rca = 2g \ .
\end{equation}
% experiment
In experiments on two coupled microring resonators the coupling strength has been measured to be $g \approx 10^{12}$s$^{-1}$~\cite{HHW17}. The response strength is therefore for this experiment $\rca \approx 2\times 10^{12}$s$^{-1}$. Note that this is several orders larger than in our example in the previous subsection. However, it is not straightforward to compare the response strength at two EPs in totally different experimental settings because of a different frequency scale and different kind of perturbations.

% not passive, zeta
One reason for a relatively large $\rca$ is that this system is not passive. Hence, the upper bound in Eq.~(\ref{eq:passiveEP2}) does not apply and $\rca$ can become arbitrarily large when the coupling strength $g$ (and accordingly the gain/loss coefficient $\alpha$) is increased.
% rcb
Correspondingly, the eigenstate response strength can become arbitrarily small, as $\rcb = 1/(2g)$, see Eq.~(\ref{eq:rcarcb}). With the coupling strength in Ref.~\cite{HHW17} we get $\rcb \approx 5\times 10^{-13}$s for this experiment.

% GF
The problem discussed in the last subsection concerning the Green's function in Eqs.~(\ref{eq:GFasym1}) and~(\ref{eq:GFasym2}) does not apply here as the imaginary part of $\evEP$ is zero and the system is not passive.

% PT-symmetric trimer
\subsection{Parity-time-symmetric trimer}
\label{sec:trimer}
In both previous examples of second-order EPs, the eigenstate response strength $\rcb$ is simply related to $\rca$ by Eq.~(\ref{eq:rcarcb}). This is different in the next example of the parity-time-symmetric trimer (see, e.g., Ref.~\cite{DG12})
\begin{equation}\label{eq:H0EPd}
\Hs = \left(\begin{array}{ccc}
\omega_0 + i\alpha & g & 0 \\
g & \omega_0 & g\\
0 & g & \omega_0 -i\alpha\\
\end{array}\right) \ .
\end{equation}
Again, $\omega_0$ is the real-valued frequency, $\alpha \geq 0$ the gain/loss coefficient, and $g \geq 0$ the coupling strength. The characteristic equation of the eigenvalue problem of $\Hs$ is 
\begin{equation}
(E_j-\omega_0)\left(\alpha^2-2g^2+(E_j-\omega_0)^2\right) = 0 \ ,
\end{equation}
with $j = 1,2,3$. Adjusting $\alpha$ to be $\sqrt{2}g$ generates a third-order degeneracy, which is an EP$_3$ as discussed for instance in the Methods section of Ref.~\cite{HHW17}.

The calculation of the response strengths in Eqs.~(\ref{eq:rca}) and~(\ref{eq:rcb}) here is more lengthy but still straightforward, one finally gets
\begin{equation}\label{eq:PTTrc}
\rca = 4g^2 
\;\;\text{and}\;\;
\rcb = \frac{1}{2g} \ .
\end{equation}
In experiments on three coupled microring resonators constituting an EP$_3$ the coupling strength has been measured to be $g \approx 9\times 10^{11}$s$^{-1}$~\cite{HHW17}. Hence, $\rca \approx 3.2\times 10^{24}$s$^{-2}$ and $\rcb \approx 5.6\times 10^{-13}$s.
We observe from Eqs.~(\ref{eq:PTTrc}) that the simple relation between the two response strengths, derived for the special case of $\order = 2$, in Eq.~(\ref{eq:rcarcb}) does not hold here. But still, a large spectral response implies a weak eigenstate response and vice versa. Note that in this example, as well as in the next one, $\rca$ and $\rcb$ are related by the trivial fact that the considered model possesses only one free relevant parameter.

% EPn in Jordan normal form
\subsection{Fully asymmetric hopping model}
\label{sec:HNmodel}
We look at a generalization of the Hamiltonian~(\ref{eq:H0EP}) 
\begin{equation}\label{eq:HNmodel}
\Hs = \left(\begin{array}{ccccc}
E_0    & A_0    & 0       & \ldots & 0  \\
0      & E_0    & A_0     & \ldots & 0  \\
0      & 0      & E_0     & \ldots & 0  \\
\vdots & \vdots & \vdots  &  \ddots       & \vdots   \\
0      & 0 & 0       &  \ldots     & E_0\\
\end{array}\right) \ .
\end{equation}
This $\order\times\order$ Hamiltonian can also be understood as the nonperiodic, fully asymmetric limiting case of the Hatano-Nelson model of a cylindrical superconductor~\cite{HN96}. This Hamiltonian describes a unidirectional hopping in a nearest-neighbor tight-binding chain with complex hopping parameter~$A_0$. For $A_0\neq 0$, the Hamiltonian is at an EP$_\order$ with eigenvalue $\evEP = E_0$.
The calculation of the response strengths is here relatively simple even for arbitrary $\order$
\begin{equation}\label{eq:HNM1}
\rca = |A_0|^{\order-1} 
\;\;\text{and}\;\;
\rcb = \frac{1}{|A_0|} \ .
\end{equation}

Solving the eigenvalue problem of the decay operator~$\oHami$ in Eq.~(\ref{eq:decayop}) for the case $\order = 3$ gives one zero eigenvalue and two further eigenvalues $-2\imagc{\evEP}\pm\sqrt{2}|A_0|$. Stipulating $\oHami$ to be positive semidefinite for a passive system leads to $\imagc{\evEP} \leq 0$ and
\begin{equation}\label{eq:HNM3}
|A_0|^2 \leq 2|\imagc{\evEP}|^2 \ .  
\end{equation}
With $\rca = |A_0|^2$ from Eq.~(\ref{eq:HNM1}) this is consistent with the upper bound in Eq.~(\ref{eq:passiveEP3}). 

\subsection{Random Hamiltonians at and near EPs}
Finally, we present an entire class of examples based on a random matrix approach. Using MATLAB we numerically construct a Hamiltonian $\Hs$ at an EP$_\order$ with eigenvalue $\evEP$ via a similarity transformation $\Hs = \op{Q}\op{J}\op{Q}^{-1}$. Here, $\op{J}$ is an $\order\times\order$ matrix at an EP$_\order$ in Jordan normal form [Eq.~(\ref{eq:HNmodel}) with $E_0 = \evEP$ and $A_0 = 1$] and $\op{Q}$ is, in general, a non-unitary, $\order\times\order$ matrix consisting of complex random numbers where real and imaginary parts are uniformly distributed in the interval $[-1/2,1/2]$. We call such a constructed $\Hs$ a \quoting{random Hamiltonian at an EP} even though its matrix elements are not completely random but have to conspire such that the Hamiltonian is at an EP.
Where needed, the perturbation $\Hp$ is chosen to be an $\order\times\order$ matrix consisting of complex random numbers with real and imaginary parts being drawn from a uniform distribution on $[-1/2,1/2]$. 

\begin{figure}[ht]
\includegraphics[width=0.95\columnwidth]{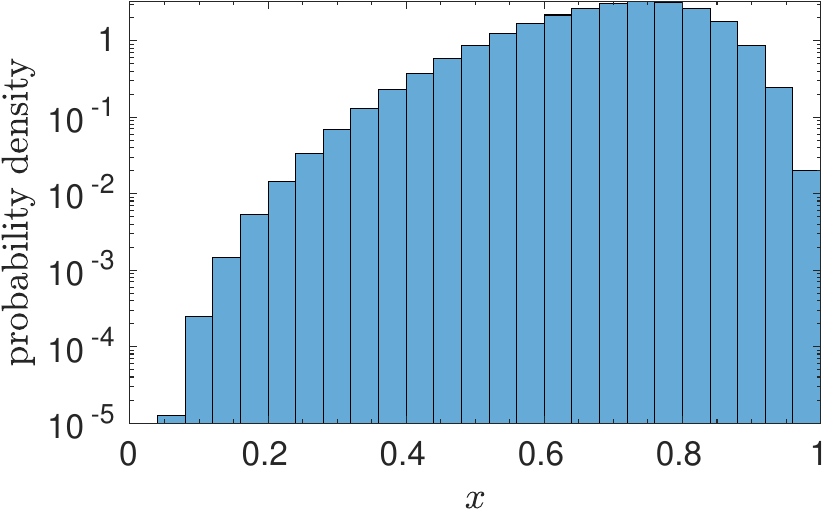}
\caption{Probability density function of the dimensionless spectral response $x$ defined in Eq.~(\ref{eq:x1}) computed from $10^7$ realizations of random Hamiltonians at an EP$_3$; see main text. The parameters are $\evEP = -i0.5$ and $\varepsilon = 10^{-7}$. Note the logarithmic scale on the vertical axis.}
\label{fig:histosplitting}
\end{figure}

% spectral response to perturbations
To study the spectral response to perturbations, we numerically compute the eigenvalues $\ev_j$ of $\Ha = \Hs+\varepsilon\Hp$ and insert them into the nonnegative quantity
\begin{equation}\label{eq:x1}
x := \frac{\text{max}(|\ev_j-\evEP|)}{(\varepsilon\normspec{\Hp}\,\rca)^{1/\order}} \ ,
\end{equation}
which, according to inequality~(\ref{eq:specresponse}), should be smaller or equal unity. Figure~\ref{fig:histosplitting} shows a histogram resulting from $10^7$ realizations of Hamiltonians in our random matrix approach for the case $\order = 3$. While the precise shape of the distribution may depend on the chosen ensemble of random Hamiltonians, it can be clearly seen that (i) $x \leq 1$ which demonstrates the validity of the inequality~(\ref{eq:specresponse}), (ii) the majority of the realizations are located well above $x = 0.5$, and (iii) the upper bound given by the response strength~$\rca$ seems to be sharp.  

\begin{figure}[ht]
\includegraphics[width=0.95\columnwidth]{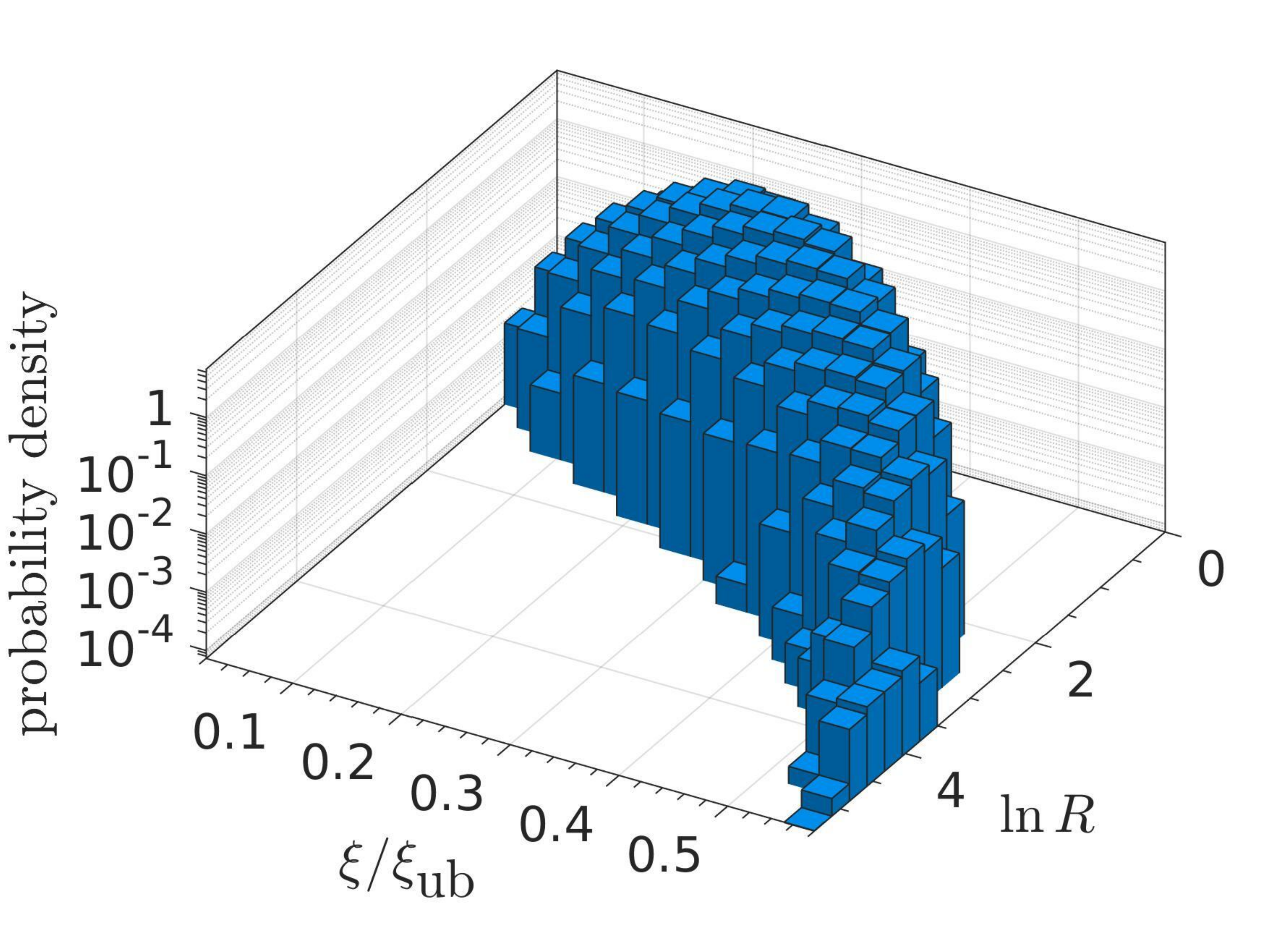}
\caption{Probability density function of the dimensionless ratio $\rca/\rcamax$ with upper bound $\rcamax$ from Eq.~(\ref{eq:passiveEP3}) and the logarithm of the dimensionless ratio $R$ of the largest and the second-largest eigenvalue of the decay operator~$\oHami$~[Eq.~(\ref{eq:decayop})]. The computation is based on $1.8\times 10^6$ realizations of random Hamiltonians at an EP$_3$ with positive semidefinite $\oHami$. The energy eigenvalue is $\evEP = -i0.5$. Note the logarithmic scale on the vertical axis.}
\label{fig:histors}
\end{figure}

% upper bound of rca for passive systems 
Next, we use the random Hamiltonians~$\Hs$ at an EP as a stress test of the upper bound $\rcamax$ of the spectral response strength~$\rca$ in passive systems. From a total of $10^9$ realizations we select those that have a positive semidefinite decay operator~$\oHami$, see Eq.~(\ref{eq:decayop}), which are around $1.8\times 10^6$. We introduce the quantity $R$ as the ratio of the largest and the second-largest eigenvalue of~$\oHami$. Large values of $R$ indicate an approximate satisfaction of $\rank{\oHami} = 1$. As argued in Sec.~\ref{sec:passive}, $\rca$ is maximized if $\rank{\oHami} = 1$. The upper bound is given  in Eq.~(\ref{eq:passiveEP2}) for $\order = 2$ and Eq.~(\ref{eq:passiveEP3}) for $\order = 3$.
Figure~\ref{fig:histors} shows a histogram of the distribution of the normalized $\rca$ and $R$ for random Hamiltonians at an EP$_3$. The numerical data clearly confirm that $\rca \leq \rcamax$ with Eq.~(\ref{eq:passiveEP3}), but we can also see that it is not a sharp bound ($\rca/\rcamax < 0.572 \approx 1/\sqrt{3}$) as suspected from the presence of the additional step based on the inequality~(\ref{eq:ineqNN}) in the derivation of the bound. Moreover, Fig.~\ref{fig:histors} reveals a correlation between $\rca$ and $R$, which indicates that if $\oHami$ approaches a rank-1 matrix for large $R$, $\rca$ exhibits its largest values, precisely confirming our expectation. 

% eigenstate response
Now, we compute numerically the eigenvectors $\ket{\state_j}$ of $\Ha = \Hs+\varepsilon\Hp$ and plug them into the nonnegative quantity
\begin{equation}\label{eq:x2}
y := \frac{\text{max}(\normV{\Delta\state_j})}{(\varepsilon\normspec{\Hp}\,\rcb)^{1/\order}} \ .
\end{equation}
Interestingly, from Eq.~(\ref{eq:evev}) together with Eq.~(\ref{eq:rcb}) follows the prediction $y = x$, i.e., the dimensionless eigenstate response equals the dimensionless spectral response in Eq.~(\ref{eq:x1}). Hence, the numerical results for $10^7$ realizations of random Hamiltonians at an EP$_3$ (not shown) look as in Fig.~\ref{fig:histosplitting}.

\begin{figure}[ht]
\includegraphics[width=0.95\columnwidth]{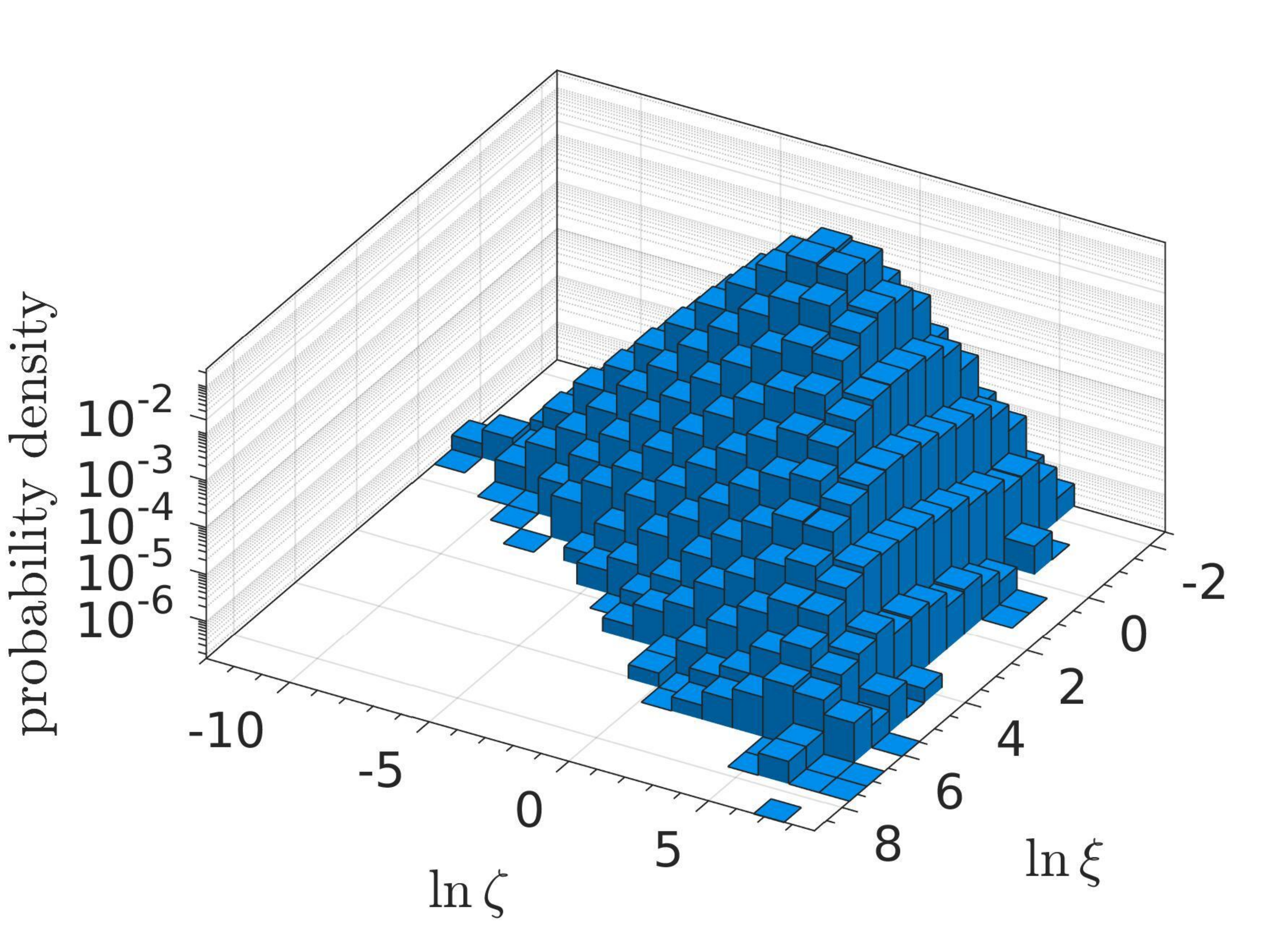}
\caption{Probability density function of the logarithm of $\rca$ ($\rca$ in units of Joule$^2$) and $\rcb$ ($\rcb$ in units of 1/Joule) computed from $10^7$ realizations of random Hamiltonians at an EP$_3$; $\evEP = -i0.5$.}
\label{fig:histocorrelation}
\end{figure}

We have seen that there is an anticorrelation of $\rca$ and $\rcb$ in the case of EP$_2$ dictated by Eq.~(\ref{eq:rcarcb}) and also for the EP$_3$ single-parameter examples in Secs.~\ref{sec:trimer} and~\ref{sec:HNmodel}. In general, however, there is no such anticorrelation. This is demonstrated in Fig.~\ref{fig:histocorrelation} with a histogram of values $\rca$ and $\rcb$ for random Hamiltonians at an EP$_3$. Hence, in general one cannot state that a large $\rca$ implies a small $\rcb$ and vice versa.

\begin{figure}[ht]
\includegraphics[width=0.95\columnwidth]{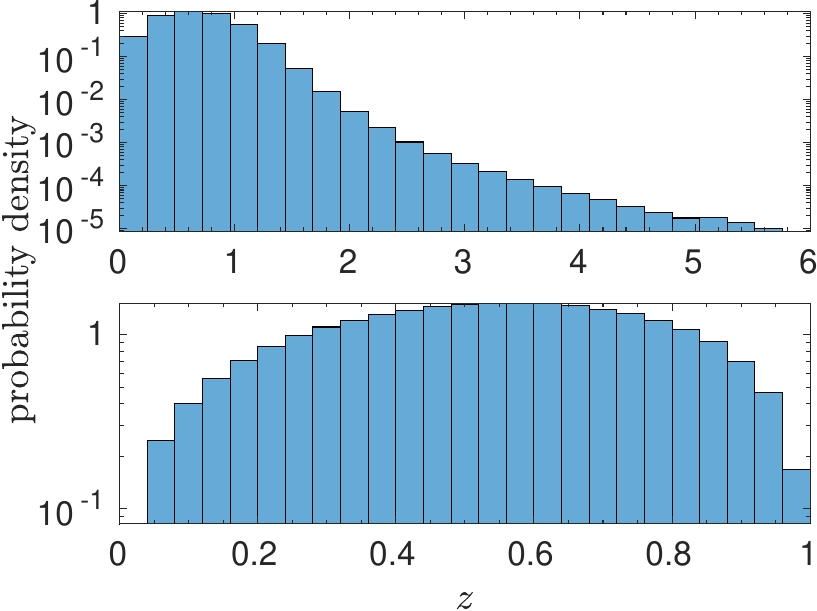}
\caption{Probability density function of the dimensionless intensity response~$z$ [Eq.~(\ref{eq:x3})] computed from $10^7$ realizations of random Hamiltonians at an EP$_3$; $\pf = 0$. The energy eigenvalue is $\evEP = -i0.5$ (upper panel, the range has been restricted to $z\leq 6$) and $\evEP = -i0.005$ (lower panel).}
\label{fig:histointensity}
\end{figure}

% intensity response
Next, we validate the upper bound for the intensity response to harmonic excitations. We compute the long-time limit $\ket{\psi}$ from Eq.~(\ref{eq:particular}) and the response strength $\rca$ from Eq.~(\ref{eq:rca}) for random Hamiltonians $\Hs$ at an EP$_\order$, random excitation vectors $\ket{p}$ normalized to unity, and given excitation frequency $\pf$ and excitation power $P$. We consider the nonnegative quantity
\begin{equation}\label{eq:x3}
z := \frac{\normV{\psi}|\hbar\pf-\evEP|^n}{P\rca} \ ,
\end{equation}
which, according to inequality~(\ref{eq:rtoe}), should be smaller or equal unity. Figure~\ref{fig:histointensity} shows a histogram resulting from $10^7$ realizations of random excitation vectors and random Hamiltonian at an EP for the case $\order = 3$. The resonant case $\hbar\pf = \realc{\evEP}$ is considered. In the upper panel we observe that for $|\hbar\pf-\evEP| = 0.5$ the upper bound is violated; around 17 percent of realizations give $z > 1$. The reason is that in the expansion of the Green's function in Eq.~(\ref{eq:Heiss}) the highest-order term is not dominant because $\hbar\pf\in\R$ does not come close to the complex eigenvalue $\evEP$. For much smaller $|\hbar\pf-\evEP|$ (the value 0.005 is chosen in the lower panel of Fig.~\ref{fig:histointensity}) the highest-order term becomes dominant and, hence, the upper bound is fulfilled, $z \leq 1$. 

% dynamics
We have also applied the random matrix approach to verify inequality~(\ref{eq:dynamics}) which quantifies the dynamic response to initial deviations from the EP eigenstate. Again, the random Hamiltonian~$\Hs$ is at an EP of order $\order = 3$. The initial state is chosen randomly and is then normalized to unity. The numerics (not shown) are fully consistent with the bound~(\ref{eq:dynamics}) and indicates that the bound is sharp.

\section{Summary}
\label{sec:summary}
We have introduced two quantities $\rca$ and $\rcb$ that characterize a general $\order\times\order$ Hamiltonian at an EP of arbitrary order $\order$. 
% rca
The spectral response strength $\rca$ quantifies not only the spectral response to perturbations but also the intensity response to excitations and the dynamic response to initial deviations from the EP eigenvector.
% rcb
The eigenstate response strength $\rcb$ quantifies the eigenstate response to perturbations.
% independce of rca and rcb
While in the special case of second-order EPs the two response strengths are related by $\rcb = 1/\rca$, no such relationship exists for higher-order EPs. Hence, both response strengths are in general independent. 

% passive systems
For passive systems $\rca$ is bounded from above. For EPs of order $\order \leq 3$ an upper bound has been derived. This bound is of particular relevance for EP-based sensors as it limits the resolvability of energy and frequency splittings.

% examples
The findings have been illustrated by a number of physically relevant examples. Moreover, the obtained bounds have been systematically tested with a random matrix approach.

% which norm
The matrix norm that enters the definition of $\rca$ and $\rcb$ has to be compatible with the vector 2-norm and unitarily invariant. Apart from these conditions the norm can be freely chosen. However, the tightest bound for the various responses is given by the spectral norm, so this is the natural choice. However, if one is interested only in the quantities $\rca$ and $\rcb$ then the Frobenius norm can be used as well as it leads to exactly the same numerical values which facilitates an easy computation. 

%%%%%%%%%%%%%%%%%%%%%%%%%%%%%%%%%%%%%%%%%%%%%%%%%%%%%%%%%%%%%%%%%%%%%%%%%%%%%%%
% applications
%%%%%%%%%%%%%%%%%%%%%%%%%%%%%%%%%%%%%%%%%%%%%%%%%%%%%%%%%%%%%%%%%%%%%%%%%%%%%%%
% sensors based on EP of higher order
The spectral response strength $\rca$ can be very beneficial for the design of higher-order EPs in particular for sensing applications. The maximal amount of energy splitting, a quantity that is essential for the sensitivity of such a device, is directly quantified by $\rca$. 
% exceptional surfaces
In this context it is interesting to combine our approach with the concept of exceptional surfaces. Each point on such a surface embedded in a higher dimensional parameter space corresponds to an EP. When the system is perturbed tangential to the surface the system stays on an EP, whereas a perturbation perpendicular to the surface leads to a deviation from the EP and therefore to an enhanced energy splitting. Exceptional surfaces have been suggested~\cite{ZRK19,ZNO19} as a possibility to remove the harmful consequences of fabrication intolerances. Here, one could use the response strength~$\rca$ to systematically search for high-response regions on the exceptional surface. 

% rcb for robust chiral states
A small eigenstate response strength $\rcb$ means that the energy eigenstate at the EP is rather robust under perturbation of the system. This kind of robustness at EPs was so far unappreciated. Without knowing one has possibly already taken advantage of it in the case of unidirectional lasing~\cite{POL16} and orbital angular momentum lasing~\cite{MZS16}. Both examples are based on an EP$_2$. The new characteristic $\rcb$ helps to study this robustness also for higher-order EPs.

\acknowledgments 
Fruitful discussions with J. Kullig are acknowledged. %We acknowledge support for the Book Processing Charge by the Open Access Publication Fund of Magdeburg University.

\begin{appendix}
\section{Green's function near an EP}
\label{app:GF}
This appendix provides a short and elegant derivation of the Green's function $\GF(\ev)$ near an EP of order~$\order$ in Eq.~(\ref{eq:Heiss}). Starting from the definition of the Green's function in Eq.~(\ref{eq:GFdef}) we write
\begin{equation}
(\ev-\evEP)\GF(\ev)-(\Hs-\evEP\ident)\GF(\ev) = \ident \ ,
\end{equation}
where $\evEP$ is the eigenvalue of $\Hs$ at the EP. We move the second term to the RHS of the equation, utilize the definition of $\op{N}$ in Eq.~(\ref{eq:N}), and divide by $\ev-\evEP$ leading to
\begin{equation}
\GF(\ev) = \frac{\ident}{\ev-\evEP} + \frac{\op{N}}{\ev-\evEP}\GF(\ev)\ . 
\end{equation}
Next, this expression is plugged into the $\GF(\ev)$ on the RHS of the equation which gives
\begin{equation}
\GF(\ev) = \frac{\ident}{\ev-\evEP} + \frac{\op{N}}{(\ev-\evEP)^2} + \frac{\op{N}^2}{(\ev-\evEP)^2}\GF(\ev)\ . 
\end{equation}
We repeat this iterative scheme until the generated series is truncated by the nilpotency of $\op{N}$ with the index of nilpotency being $\order$. With definition~(\ref{eq:Mk}) one obtains the result in Eq.~(\ref{eq:Heiss}).

%\section{Derivation of inequality~(\ref{eq:ineqNN})}
%\label{app:ineqNN}

\section{Relation between $\normM{\Gn_\order}$ and the norm of the Jordan vector $\ket{\Jordan_n}$}
\label{app:Jordan}
In this appendix we derive a relation between $\normFro{\Gn_\order}$ and the norm of the Jordan vector $\ket{\Jordan_n}$. From Eqs.~(\ref{eq:jc})  and~(\ref{eq:N}) follows $\op{N}^{\order-1}\ket{\Jordan_n} = \ket{\Jordan_1}$ which together with the definition of $\Gn_k$ in Eq.~(\ref{eq:Mk}) leads to 
\begin{equation}
\sandwich{\Jordan_n}{\Gn_\order^\dagger\Gn_\order}{\Jordan_n} = \braket{\Jordan_1}{\Jordan_1} \ .
\end{equation}
With the normalization~(\ref{eq:ortho1}) the RHS of this equation is unity. With the orthogonalization~(\ref{eq:ortho2}) the LHS can be written as trace over $\Gn_\order^\dagger\Gn_\order$ using an orthonormal basis with one element being the unit vector
\begin{equation}
\ket{u_\order} = \frac{\ket{\Jordan_\order}}{\normV{\Jordan_\order}} \ .
\end{equation}
Recalling the definition of the Frobenius norm~(\ref{eq:Fronorm}) the result finally is
\begin{equation}\label{eq:rcaJv}
\normFro{\Gn_\order} = \frac{1}{\normV{\Jordan_\order}} \ .
\end{equation}
It is to emphasize that this is true only if the normalization and orthogonalization conditions~(\ref{eq:ortho1})-(\ref{eq:ortho2}) are employed.

\end{appendix}

%\bibliography{../../../bib/fg4,../../../bib/extern}
%apsrev4-2.bst 2019-01-14 (MD) hand-edited version of apsrev4-1.bst
%Control: key (0)
%Control: author (8) initials jnrlst
%Control: editor formatted (1) identically to author
%Control: production of article title (0) allowed
%Control: page (0) single
%Control: year (1) truncated
%Control: production of eprint (0) enabled
%

\end{document}